%% file: ProjectStructure.tex
\documentclass[conference]{IEEEtran}

\usepackage{amsmath,amssymb,amsfonts,latexsym}
\usepackage{enumerate}
\usepackage{xspace}
\usepackage{epsf,picinpar}
\usepackage{varioref}
\usepackage{colortbl,multirow,hhline}
\usepackage{listings}
\usepackage{amssymb}
\usepackage{comment}

\usepackage{algorithmic}
\usepackage{algorithm}
\usepackage{caption}
\usepackage[normalem]{ulem}
\usepackage{pifont}
\usepackage{url}
\usepackage{balance}
\usepackage{graphicx, subfigure}
\usepackage{longtable}
\usepackage{lscape}
\usepackage{multirow}
\usepackage{listings}
\usepackage{framed}
\usepackage{morefloats}
\usepackage[T1]{fontenc}
\usepackage{array}
\usepackage{pdfpages}
\usepackage{fancybox}
\usepackage{amsmath}
\usepackage{flushend}
\usepackage{float}
\usepackage{booktabs}
\usepackage{enumitem}
\usepackage{tabularx}
\usepackage{longtable}
\usepackage{placeins}

\usepackage{arydshln}
\definecolor{goodgreen}{RGB}{198,239,206}
\definecolor{badred}{RGB}{255,199,206}

\begin{document}

\title{\fontsize{23pt}{28pt}\selectfont An Empirical Study on How Architectural Topology Affects Microservice Performance and Energy Usage}


\author{
Irena Ristova,
Vincenzo Stoico \\
Vrije Universiteit Amsterdam, Amsterdam, The Netherlands\\
Email: i2.ristova@student.vu.nl, v.stoico@vu.nl \\
}
\maketitle

\begin{abstract}
Microservice architectures form the backbone of modern software systems for their scalability, resilience, and maintainability, but their rise in cloud-native environments raises energy efficiency concerns.
While prior research addresses microservice decomposition and placement, the impact of topology, the structural arrangement and interaction pattern among services, on energy efficiency remains largely underexplored.
This study quantifies the impact of topologies on energy efficiency and performance across six canonical ones (Sequential Fan-Out, Parallel Fan-Out, Chain, Hierarchical, Probabilistic, Mesh), each instantiated at 5-, 10-, and 20-service scales using the $\mu\text{Bench}$ framework. We measure throughput, response time, energy usage, CPU utilization, and failure rates under an identical workload. 
The results indicate that topology influences the energy efficiency of microservices under the studied conditions. As system size increases, energy consumption grows, with the steepest rise observed in dense Mesh and Chain topologies. 
Mesh topologies perform worst overall, with low throughput, long response times, and high failure rates. Hierarchical, Chain, and Fan-Out designs balance performance and energy use better. As systems scale, metrics converge, with Probabilistic and Parallel Fan-Out emerging as the most energy-efficient under CPU-bound loads. These results guide greener microservice architecture design and serve as a baseline for future research on workload and deployment impacts.


\end{abstract}


\input{1.intro}

\input{2.related}

\input{3.definition}
\input{4.planning} 
\input{5.execution}

\input{6.results}
\input{7.discussion}

\input{8.threats}
\input{9.conclusion} 
\vspace{-.1cm}
\section*{Acknowledgements}
We would like to thank Dr. Ivano Malavolta for his support, feedback, and helpful comments on this work.

\bibliographystyle{IEEEtran}
\bibliography{references}
\clearpage


\end{document}

%% file: 1.intro.tex
\section{Introduction}

Microservices have transformed modern software design, deployment, and maintenance. The concept gained traction after the influential 2014 article by Martin Fowler and James Lewis \cite{lewis_fowler_2014}. This style structures an application as a set of small, independent services communicating through lightweight mechanisms, typically HTTP-based APIs \cite{newman2021building}. Today, companies such as Netflix, Uber, and Amazon use microservices to achieve scalability, resilience, and maintainability, making it the standard for large-scale systems.
%
%

Literature shows that microservices can significantly improve software quality across various dimensions \cite{avritzer2025architecture, rodrigues2025assessment, ozdemir2024evaluating, ahmad2024smart}.
Yet as adoption increases, energy usage has become a major concern \cite{xiao2025effectiveness, floroiu2024anomaly, werner2025comprehensive}. Greater workloads often require horizontal scaling and more service replicas, which amplify communication overhead and resource inefficiency \cite{legler2025service}.
Architectural choices strongly influence energy consumption: Zhao et al. \cite{zhao2025does} find that finer-grained designs can use up to 13\% more energy, while Capra et al. \cite{capra2012measuring} show that simpler, less layered architectures tend to be more energy-efficient.

Microservices follow common architectural patterns, such as the database-per-service model and API gateways, to define the structure and deployment of services \cite{richardson2018microservices}. This overall structure, referred to as the \textbf{topology} of the application \cite{cervantes2024designing}, determines system complexity and can significantly affect energy efficiency.


The \textbf{goal} of this study is to determine whether the topology of a microservice-based application influences its performance and energy efficiency.
Our study is guided by the following research questions:

\noindent \textbf{$\mathbf{RQ_1}$} --\textit{ What is the impact of the architectural topology on the energy efficiency of microservices?}

\noindent \textbf{$\mathbf{RQ_2}$} -- \textit{How does performance correlate to energy usage across topologies?}

We conducted an experiment comparing six microservice topologies, Sequential Fan-Out, Parallel Fan-Out, Chain, Hierarchical, Probabilistic, and Mesh, implemented with 5, 10, and 20 services. Each topology was evaluated under identical workloads for energy usage and performance, using $\mu\text{Bench}$ \cite{detti2023mubench}, a framework for generating synthetic microservice systems.
A major challenge in studying energy efficiency at the architectural level is the absence of reusable benchmarking datasets \cite{pham2024root}. Existing benchmarks target fixed applications, limiting analysis of architectural factors such as service dependencies and concurrency. To address this, we used $\mu\text{Bench}$ to generate synthetic applications implementing each topology.


%

Results indicate that topology strongly influences energy–performance behavior. Dense Mesh and Chain topologies show higher energy use (up to 39.13 kJ at 20 services) due to coordination overhead, while Probabilistic routing is most efficient at smaller scales ($\approx$ 29 kJ at 5–10 services). Energy stabilizes around 38–39 kJ as systems grow. Trade-offs vary by structure: Hierarchical and fan-out topologies link higher performance with lower energy, whereas Mesh reverses this pattern.
The main \textbf{contributions} are: (1) six $\mu\text{Bench}$-based topologies, (2) an experiment comparing their energy–performance impact, and (3) a replication package \cite{ReplicationPackage} with scripts, data, and topology models. These findings support the design of greener microservice systems and supply reusable data for energy–performance research.

%% file: 2.related.tex
\section{Related Work}
\label{sec:related}

Empirical research on microservice architectures has primarily focused on performance, scalability, and reliability~\cite{pahl2016microservices,kaushik2023empirical}. Existing work spans three main areas: benchmarking frameworks, empirical dependency analysis, and topology-based architectural studies.

$\mu$Bench, proposed by Detti et al.~\cite{detti2023mubench}, is a configurable framework for generating Kubernetes-based microservice benchmarks with controlled dependencies and workloads. It supports experiments on structural features like centralized vs. distributed meshes and sequential vs. parallel patterns, showing that topology strongly affects throughput and latency, especially in centralized and fan-out setups.

Large-scale trace analyses confirm that call-graph structure critically affects system performance. Studies of Alibaba Cloud and other production systems show strong correlations between fan-in/out, dependency depth, and graph complexity with performance variability~\cite{zhang2019,luo2021,du2024}. Recent work on microservice observability and topology-aware scheduling~\cite{pham2024root,hao2022,li2023topology} further underscores that inter-service dependencies and architectural topology are major factors influencing runtime behavior, throughput, and latency.


Research on structural complexity supports this view. Van der Laan~\cite{vanderlaan2023} shows that topology metrics indicate architectural quality, while Bakhtin et al.~\cite{bakhtin2025network} link high service centrality to coordination bottlenecks. Although performance effects are well studied, energy use in microservice remains underexplored~\cite{araujo2024energyMicroservices}. Recent experiments reveal that architectural tactics influence both energy and performance~\cite{xiao2025effectiveness}, and energy efficiency is increasingly treated as a core architectural concern~\cite{pahl2021energyTactics}. This study extends prior work by integrating system-level energy measurements into topology-aware benchmarking with predefined $ \mu\text{Bench} $ workmodels to isolate communication structure effects.

%% file: 4.planning.tex
\section{Subjects Selection}
\label{sec:experimental_subjects_and_topology_families}

Table \ref{tab:topologies} summarizes the selected microservice topologies derived from $\mu\text{Bench}$ workmodels. Each workmodel defines a distinct service interaction pattern that determines communication and operation in virtualized environments such as Kubernetes. By analyzing service call graphs, these workmodels can be seen as instances of specific microservice topologies. To ensure validity, we include only topologies already implemented in $\mu\text{Bench}$.

\begin{table}
    \centering
    \footnotesize
    \caption{Selected Microservice Topologies.}
    \label{tab:topologies}
    \begin{tabularx}{\linewidth}{lX}
        \toprule
        \textbf{Topology Name} & \textbf{Structural Description} \\
        \midrule
        Sequential Fan-Out &
        Centralized hub-and-spoke structure with strictly sequential invocation of downstream services. \\\hdashline
        Parallel Fan-Out &
        Centralized hub-and-spoke structure with concurrent invocation of downstream services. \\\hdashline
        Chain with Branching &
        Linear sequence of services with forward branches creating alternative execution paths. \\\hdashline
        Hierarchical Tree &
        Balanced tree structure with deterministic parent--child service invocations. \\\hdashline
        Probabilistic Tree &
        Tree structure with probabilistic service invocations and variable execution paths. \\\hdashline
        Complex Mesh &
        Densely connected service network without centralized control. \\
        \bottomrule
    \end{tabularx}
    \vspace{-0.5cm}
\end{table}

Across all topology families, each service executes the \textit{same} internal workload when handling a request. The workload uses the $\mu\text{Bench}$ \emph{loader} configured for CPU stress only: computing $\pi$ to 100 decimal places and repeating this computation 10--20 times per request in a single thread. Memory, disk, and sleep stress are disabled. Thus, observed differences in performance and energy consumption stem from communication structure and invocation patterns rather than per-service computation.

To facilitate comparison, we assign descriptive names to each topology based on their structural features \cite{detti2023mubench}.
Rather than focusing on application-specific architectures, the selection targets fundamental communication structures that differ in key architectural dimensions. These dimensions include \textit{centralization}, \textit{fan-in} and \textit{fan-out} distribution, \textit{execution order} (sequential versus parallel), \textit{branching behavior}, \textit{dependency path length}, and \textit{connectivity density}. Graph-based properties such as fan-in and fan-out are widely used to characterize microservice dependency structures and architectural complexity~\cite{vanderlaan2023,luo2021,bakhtin2025network}. Fan-out refers to the number of outgoing service calls initiated by a service, while fan-in refers to the number of incoming calls received. 

The selected topology families span a diverse range of interaction patterns commonly observed in microservice systems. These include centralized orchestration patterns resembling API Gateway architectures~\cite{richardson2020}, request-processing pipelines~\cite{hao2022}, hierarchical decompositions and layered designs~\cite{vanderlaan2023,li2023topology}, conditional routing structures~\cite{luo2021}, and densely connected service meshes frequently associated with organically evolved systems~\cite{du2024, bakhtin2025network, taibi2017microservices}. 

The \textit{Sequential Fan-Out} topology represents a centralized hub-and-spoke structure with strictly sequential execution semantics, in which a single entry-point service invokes downstream services one after another. Although the communication graph resembles a star, the execution path is linear, reflecting orchestrated service pipelines~\cite{hao2022}.
The \textit{Parallel Fan-Out} topology retains the centralized structure but introduces concurrency, with the entry-point service invoking multiple downstream services simultaneously, corresponding to parallelized gateway-driven architectures~\cite{richardson2020}.
The \textit{Chain with Branching} topology extends a linear chain with forward branches, introducing alternative paths yet preserving a main execution flow~\cite{hao2022}.
The \textit{Hierarchical Tree} topology forms layered parent–child dependencies reflecting structured service hierarchies~\cite{vanderlaan2023,li2023topology}.
The \textit{Probabilistic Tree} topology adds branch probabilities to model conditional routing and heterogeneous behavior~\cite{luo2021}.
The \textit{Complex Mesh} topology is a densely connected graph with overlapping paths and high fan-in/out, typifying tightly coupled microservice ecosystems~\cite{du2024,bakhtin2025network,taibi2017microservices}.
The $\mu\text{Bench}$ workmodel of each topology can be found in our replication package \cite{ReplicationPackage}.

\section{Experimental Variables}
\label{sec:experimental_variables}
The \textbf{independent} variables of our experiment are the \textit{topology}, that is, the structure of service-to-service interactions, generated through $\mu\text{Bench}$ workmodels, and the \textit{system size}, defined as the number of services in a deployment. Each workmodel induces a distinct communication graph characterized by variations in centralization, branching, parallel request execution, and connectivity. Three system sizes are considered: 5, 10, and 20 services. System size is used to study how architectural effects scale with increasing complexity.
As \textbf{dependent} variables, we measure \textit{energy usage} of each topology as the total energy consumed by the CPU package during workload execution in kiloJoules. For RQ2, we profile also performance metrics, such as \textit{throughput} as requests per second (RPS), defined as the number of successfully processed requests per unit time; the \textit{average response time} in seconds (s), representing the mean end-to-end latency; \textit{CPU usage}, defined as the average number of CPU cores used during execution (from container metrics; 1.0 denotes one core fully utilized), and failure rate, as percentage (\%) of failed requests.
To ensure fair and comparable results, the experiment controls several factors: the \textit{workload configuration} (keeping request rate, concurrency, and duration constant), the \textit{service configuration} (using identical $\mu\text{Bench}$ workload parameters and resource settings), the \textit{execution environment} (maintaining the same hardware, operating system, container runtime, and Kubernetes cluster), and the \textit{benchmarking procedure} (following identical deployment, measurement, and cleanup steps). By controlling these factors, any observed differences in performance or energy-related behavior can be attributed primarily to the independent variables.

\section{Experiment Design}
\label{sec:experiment_design}
The experiment follows a \textit{full factorial design} \cite{wohlin12} defined by two independent variables: communication topology and system size. Communication topology is evaluated at six levels corresponding to the selected topology families. System size is evaluated at three levels: 5, 10, and 20 services. Together, these factors define 18 unique topology–size configurations. To enable statistical comparison across configurations, each topology–size combination is executed with multiple repetitions. Following an initial validation phase consisting of one run per configuration (18 runs total) to verify correctness and end-to-end integration, the final experimental design includes \textit{ten} repetitions per configuration. This results in a total of \textbf{180 experimental runs}. Configuration order is randomized to reduce the influence of topology characteristics on subsequent runs. 

\section{Data Analysis}
\label{sec:data_analysis_plan}

The data analysis follows established principles for controlled experiments in software engineering~\cite{wohlin12}, emphasizing systematic comparison across factor levels and statistical validation of observed effects. For each dependent variable, descriptive statistics summarize central tendency and dispersion across topology–size configurations.
Beyond individual topologies, we compare groups of related structures, for example, dense vs. hierarchical or deterministic vs. probabilistic designs, to assess whether broader architectural traits influence performance and energy use. Grouping by structural families enables a generalized view of coordination and dependency patterns.
Normality is tested using the Shapiro–Wilk test. As some family-level data deviate from normality, the Kruskal–Wallis test assesses overall differences, followed by pairwise Mann–Whitney U tests with Holm correction when significant effects arise. All tests use a significance level of $\alpha = 0.05$.

%% file: 5.execution.tex
\section{Experiment Execution}
\label{sec:experiment_execution}
\begin{figure}
    \centering
    \includegraphics[width=\linewidth]{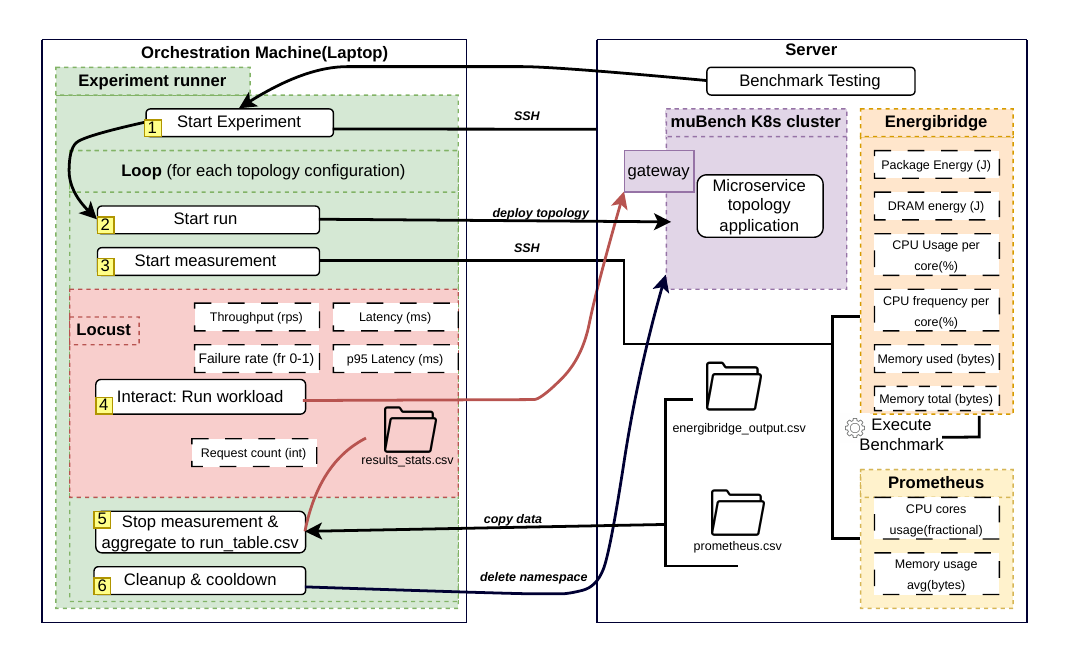}
    \caption{Experiment Execution}
    \label{fig:expflow}
    \vspace{-0.5cm}
\end{figure}

Our experimental setting is shown in Figure \ref{fig:expflow} and includes two machines: an \textit{orchestration} machine and a \textit{server}.
The former is used for experiment orchestration, workload generation, and data aggregation. It is an HP EliteBook~650 running Ubuntu~24.04.3~LTS, equipped with 16~GB RAM and a 512~GB SSD. It executes the Experiment Runner framework \cite{karsten2025experiment} to coordinate the runs, start and collect measurements, while the Locust load generator \cite{locust} generates the workload. We deploy the microservice systems generated with $\mu\text{Bench}$ on server. The server is equipped with two Intel Xeon processors (32~vCPUs in total), 384~GB RAM, and 36~TB of storage, and runs Ubuntu Server 24.04. A single-node Kubernetes cluster is instantiated using Minikube \cite{minikube} on this server. Minikube is a tool for running a single-node Kubernetes cluster locally. The cluster is allocated 28~vCPUs and approximately 350~GB of memory, leaving sufficient resources for the host system and monitoring tools to ensure stable operation during benchmarking.

All interactions between the orchestration and server machines are explicitly controlled. HTTP requests generated by Locust on the orchestrator are forwarded through SSH tunnels to the server, where they are routed to the NGINX gateway deployed as part of the $\mu\text{Bench}$ application. The gateway acts as the single entry point for all requests and dispatches them to the internal microservices according to the deployed topology. Monitoring data is accessed through a separate SSH tunnel that exposes the Prometheus API to the host machine. This setup enables secure access to runtime metrics without exposing services publicly.

\subsubsection{Toolchain}
\textit{$\mu\text{Bench}$} is used to generate and deploy synthetic microservice-based systems based on predefined workmodels. The K8sDeployer component translates workmodel specifications into Kubernetes resources. Each experimental run deploys a fresh $\mu\text{Bench}$ application instance in an isolated Kubernetes namespace. \textit{Experiment Runner} is a Python-based framework used to orchestrate the full experimental workflow \cite{karsten2025experiment}. It manages deployment, controls measurement phases, queries monitoring tools, and aggregates all collected metrics into a structured run table. \textit{Locust} is used as the workload generator. For each run, Locust executes a headless load test with 100 concurrent users over a fixed duration of 10 minutes, sending HTTP requests to the system entry point exposed by the gateway. 
Performance metrics (throughput, latency, failure rate) are collected. \textit{Prometheus} monitors pod-level resource usage within the Kubernetes cluster \cite{prometheus}, and the Experiment Runner retrieves CPU and memory data via its API. System-wide energy consumption is measured using \textit{EnergiBridge} \cite{sallou2023energibridge}, which leverages Intel RAPL counters to record CPU and DRAM energy during each run.

\subsubsection{Execution Flow}
Each experimental run follows a standardized and fully automated protocol controlled by the Experiment Runner, as illustrated in Figure~\ref{fig:expflow}. The protocol follows established experimental procedures commonly used in software energy optimization research \cite{malavolta2024ten}. During the \textit{Experiment Initialization} phase, the Experiment Runner sets up the necessary SSH connections, verifies monitoring availability, and prepares the environment for each configuration consisting of topology, size, and repetition. In the \textit{Deployment Phase}, a new Kubernetes namespace following the pattern \texttt{mubench-\{topology\}-\{size\}-\{replicate\}} is created, and the selected $\mu\text{Bench}$ workmodel is deployed using the K8sDeployer, with execution continuing only once all pods are ready and the gateway is responsive. The \textit{Measurement and Workload Phase} begins by activating EnergiBridge for energy monitoring. Locust then generates workload traffic using 100 concurrent users (spawn rate 50/s) for 10 minutes. Users repeatedly send HTTP GET requests to the ingress endpoint (\texttt{/s0}), thereby propagating requests through the deployed topology. Timestamp synchronization ensures alignment between workload execution and energy measurements. Following execution, the \textit{Metric Collection and Aggregation} phase retrieves energy readings from EnergiBridge, gathers performance data from Locust, and queries Prometheus for resource utilization metrics, which are then parsed and consolidated into a structured results table (\texttt{run\_table.csv}). Finally, during the \textit{Cleanup and Cooldown} phase, the created namespace is deleted to preserve run isolation. A 60‑second cooldown period then follows before initiating the next configuration. Because the execution of one run can affect subsequent runs through residual energy buildup, the cooldown period allows the testbed to return to its initial state, ensuring that all runs are performed under similar conditions. All configurations are executed in randomized order. Timeout mechanisms and gateway readiness checks are enforced before measurements start; runs that fail to deploy or stabilize within predefined limits are logged and excluded from analysis.

%% file: 6.results.tex
\section{Results}
\textit{Descriptive Statistics.} Table~\ref{tab:descriptive-topology} reports descriptive statistics for energy and CPU usage, throughput, response time, and failure rate. across the evaluated topologies.
For each metric, mean, standard deviation, median (50\%), minimum, maximum, and coefficient of variation (CV) are provided.

\begin{table*}
\centering
\caption{Descriptive statistics by topology and system size. Green highlights the best value for each size and metric across topologies, while red marks the worst.}
\label{tab:descriptive-topology}
\scriptsize
\setlength\tabcolsep{3pt}
\renewcommand\arraystretch{1.03}
\begin{tabular}{cl|ccc|ccc|ccc|ccc|ccc|ccc}
\toprule
& & \multicolumn{3}{c|}{\textbf{Sequential Fan-Out}} & \multicolumn{3}{c|}{\textbf{Parallel Fan-Out}} & \multicolumn{3}{c|}{\textbf{Chain}} & \multicolumn{3}{c|}{\textbf{Hierarchical}} & \multicolumn{3}{c|}{\textbf{Probabilistic}} & \multicolumn{3}{c}{\textbf{Mesh}} \\
\cmidrule(lr){3-5} \cmidrule(lr){6-8} \cmidrule(lr){9-11} \cmidrule(lr){12-14} \cmidrule(lr){15-17} \cmidrule(lr){18-20}
& & 5 & 10 & 20 & 5 & 10 & 20 & 5 & 10 & 20 & 5 & 10 & 20 & 5 & 10 & 20 & 5 & 10 & 20 \\
\midrule
\multirow{6}{*}{\textbf{\shortstack{Energy Usage \\ (kJ)}}} & \textbf{Mean} & 28.34 & 37.86 & 38.89 & 30.44 & 38.04 & 36.85 & 30.4 & 38.44 & 39.08 & 28.88 & 37.9 & 39.12 & 19.74 & 29.11 & 38.67 & 18.42 & 29.83 & 39.12 \\
 & \textbf{Min} & 28.08 & 37.72 & 38.8 & 30.08 & 37.96 & 34.98 & 30.14 & 38.4 & 38.98 & 28.64 & 37.79 & 39.05 & 19.55 & 28.9 & 38.62 & 18.28 & 29.37 & 39.09 \\
 & \textbf{50\%} & 28.35 & 37.87 & 38.89 & \cellcolor{badred} 30.37 & 38.04 & 37.31 & 30.33 & \cellcolor{badred} 38.43 & 39.09 & 28.95 & 37.92 & 39.11 & 19.71 & \cellcolor{goodgreen} 29.05 & \cellcolor{goodgreen} 38.68 & \cellcolor{goodgreen} 18.38 & 29.85 & \cellcolor{badred} 39.13 \\
 & \textbf{Max} & 28.65 & 37.95 & 39.0 & 30.79 & 38.14 & 37.91 & 30.79 & 38.5 & 39.22 & 29.11 & 37.96 & 39.18 & 19.96 & 29.42 & 38.72 & 18.84 & 30.28 & 39.19 \\
 & \textbf{Std} & 0.2 & 0.07 & 0.06 & 0.27 & 0.06 & 1.09 & 0.22 & 0.04 & 0.07 & 0.18 & 0.05 & 0.04 & 0.11 & 0.18 & 0.03 & 0.16 & 0.34 & 0.03 \\
 & \textbf{CV} & 0.71 & 0.2 & 0.14 & 0.9 & 0.16 & 2.95 & 0.72 & 0.1 & 0.18 & 0.62 & 0.14 & 0.1 & 0.54 & 0.63 & 0.08 & 0.86 & 1.14 & 0.08 \\
 \midrule
 \multirow{6}{*}{\textbf{\shortstack{CPU Usage \\ (cores)}}} & \textbf{Mean} & 1.79 & 1.57 & 0.94 & 1.87 & 1.65 & 0.88 & 1.86 & 1.72 & 0.97 & 1.81 & 1.57 & 0.97 & 1.2 & 1.03 & 0.91 & 0.88 & 1.04 & 0.97 \\
 & \textbf{Min} & 1.77 & 1.54 & 0.92 & 1.85 & 1.62 & 0.8 & 1.83 & 1.7 & 0.96 & 1.75 & 1.53 & 0.96 & 1.18 & 1.01 & 0.9 & 0.86 & 1.03 & 0.96 \\
 & \textbf{50\%} & 1.78 & 1.57 & 0.93 & \cellcolor{badred} 1.87 & 1.65 & \cellcolor{goodgreen} 0.89 & 1.87 & \cellcolor{badred} 1.72 & 0.97 & 1.82 & 1.58 & \cellcolor{badred} 0.98 & 1.2 & \cellcolor{goodgreen} 1.03 & 0.91 & \cellcolor{goodgreen} 0.88 & 1.04 & 0.97 \\
 & \textbf{Max} & 1.83 & 1.59 & 0.95 & 1.91 & 1.67 & 0.91 & 1.89 & 1.73 & 0.98 & 1.83 & 1.6 & 0.98 & 1.24 & 1.04 & 0.92 & 0.9 & 1.06 & 0.98 \\
 & \textbf{Std} & 0.02 & 0.01 & 0.01 & 0.02 & 0.02 & 0.04 & 0.02 & 0.01 & 0.01 & 0.03 & 0.02 & 0.01 & 0.02 & 0.01 & 0.01 & 0.01 & 0.01 & 0.01 \\
 & \textbf{CV} & 1.11 & 0.93 & 0.94 & 0.89 & 0.99 & 4.6 & 1.03 & 0.51 & 0.66 & 1.6 & 1.51 & 0.98 & 1.8 & 0.92 & 0.91 & 1.41 & 1.25 & 0.9 \\
 \midrule
\multirow{6}{*}{\textbf{\shortstack{Throughput \\ (rps)}}} & \textbf{Mean} & 22.69 & 17.15 & 8.58 & 24.73 & 17.1 & 7.91 & 24.65 & 17.42 & 8.8 & 23.17 & 17.19 & 8.82 & 18.29 & 23.63 & 14.72 & 6.36 & 8.6 & 8.82 \\
 & \textbf{Min} & 22.46 & 17.0 & 8.56 & 24.24 & 17.03 & 7.27 & 24.53 & 17.31 & 8.77 & 22.79 & 17.14 & 8.78 & 17.94 & 23.45 & 14.65 & 6.16 & 8.44 & 8.79 \\
 & \textbf{50\%} & 22.75 & 17.16 & 8.59 & \cellcolor{goodgreen} 24.67 & 17.11 & \cellcolor{badred} 8.04 & 24.55 & 17.44 & 8.79 & 23.19 & 17.19 & 8.82 & 18.25 & \cellcolor{goodgreen} 23.56 & \cellcolor{goodgreen} 14.72 & \cellcolor{badred} 6.4 & \cellcolor{badred} 8.6 & 8.82 \\
 & \textbf{Max} & 22.88 & 17.22 & 8.62 & 25.1 & 17.16 & 8.29 & 24.95 & 17.48 & 8.82 & 23.42 & 17.25 & 8.85 & 18.73 & 23.9 & 14.8 & 6.5 & 8.82 & 8.85 \\
 & \textbf{Std} & 0.15 & 0.06 & 0.02 & 0.31 & 0.04 & 0.37 & 0.17 & 0.06 & 0.02 & 0.19 & 0.04 & 0.02 & 0.28 & 0.15 & 0.04 & 0.13 & 0.13 & 0.02 \\
 & \textbf{CV} & 0.66 & 0.36 & 0.24 & 1.24 & 0.25 & 4.68 & 0.69 & 0.32 & 0.22 & 0.8 & 0.22 & 0.25 & 1.53 & 0.64 & 0.27 & 2.11 & 1.47 & 0.23 \\
\midrule
\multirow{6}{*}{\textbf{\shortstack{Response Time \\ (s)}}} & \textbf{Mean} & 2.4 & 3.82 & 9.56 & 2.04 & 3.83 & 10.53 & 2.05 & 3.72 & 9.28 & 2.31 & 3.8 & 9.26 & 3.46 & 2.23 & 4.77 & 13.51 & 9.52 & 9.25 \\
 & \textbf{Min} & 2.37 & 3.79 & 9.53 & 1.98 & 3.81 & 9.98 & 1.99 & 3.7 & 9.25 & 2.26 & 3.77 & 9.21 & 3.33 & 2.18 & 4.73 & 13.15 & 9.26 & 9.21 \\
 & \textbf{50\%} & 2.39 & 3.81 & 9.57 & \cellcolor{goodgreen} 2.05 & 3.82 & \cellcolor{badred} 10.29 & 2.07 & 3.72 & 9.28 & 2.31 & 3.8 & 9.26 & 3.47 & \cellcolor{goodgreen} 2.24 & \cellcolor{goodgreen} 4.77 & \cellcolor{badred} 13.43 & \cellcolor{badred} 9.51 & 9.24 \\
 & \textbf{Max} & 2.45 & 3.86 & 9.59 & 2.12 & 3.87 & 11.58 & 2.08 & 3.75 & 9.31 & 2.37 & 3.82 & 9.3 & 3.56 & 2.25 & 4.8 & 14.0 & 9.72 & 9.29 \\
 & \textbf{Std} & 0.03 & 0.02 & 0.02 & 0.05 & 0.02 & 0.61 & 0.03 & 0.02 & 0.02 & 0.03 & 0.02 & 0.03 & 0.08 & 0.03 & 0.02 & 0.32 & 0.16 & 0.03 \\
 & \textbf{CV} & 1.2 & 0.54 & 0.26 & 2.34 & 0.55 & 5.79 & 1.53 & 0.51 & 0.24 & 1.44 & 0.5 & 0.36 & 2.39 & 1.3 & 0.44 & 2.35 & 1.69 & 0.33 \\
\midrule
\multirow{6}{*}{\textbf{\shortstack{Failure rate \\ (\%)}}} & \textbf{Mean} & 0.37 & 0.38 & 0.0 & 0.43 & 0.67 & 0.0 & 0.52 & 0.69 & 2.13 & 0.39 & 0.45 & 0.72 & 0.38 & 0.23 & 0.48 & 3.14 & 2.5 & 0.71 \\
 & \textbf{Min} & 0.3 & 0.33 & 0.0 & 0.37 & 0.46 & 0.0 & 0.45 & 0.56 & 1.92 & 0.35 & 0.38 & 0.57 & 0.26 & 0.18 & 0.35 & 1.82 & 2.23 & 0.59 \\
 & \textbf{50\%} & \cellcolor{goodgreen} 0.37 & 0.38 & \cellcolor{goodgreen} 0.0 & 0.44 & 0.67 & 0.0 & 0.51 & 0.69 & \cellcolor{badred} 2.13 & 0.38 & 0.43 & 0.68 & 0.38 & \cellcolor{goodgreen} 0.22 & 0.47 & \cellcolor{badred} 2.61 & \cellcolor{badred} 2.48 & 0.66 \\
 & \textbf{Max} & 0.47 & 0.45 & 0.02 & 0.5 & 0.83 & 0.02 & 0.6 & 0.88 & 2.49 & 0.48 & 0.62 & 0.98 & 0.5 & 0.28 & 0.64 & 5.19 & 2.88 & 0.91 \\
 & \textbf{Std} & 0.05 & 0.03 & 0.01 & 0.06 & 0.11 & 0.01 & 0.04 & 0.1 & 0.21 & 0.04 & 0.07 & 0.11 & 0.06 & 0.04 & 0.09 & 1.16 & 0.23 & 0.12 \\
 & \textbf{CV} & 13.62 & 8.3 & 316.23 & 12.97 & 16.07 & 210.82 & 8.08 & 13.91 & 9.9 & 9.73 & 15.23 & 15.62 & 15.11 & 17.69 & 18.57 & 36.89 & 9.17 & 16.28 \\
\bottomrule
\end{tabular}
\vspace{-0.5cm}
\end{table*}

Energy usage increases consistently with system size across all topologies. At 5 services, Mesh shows the lowest median energy of 18.38 kJ, while Parallel Fan-Out has the highest at 30.37 kJ. As system size reaches 10 services, medians range from 29.05 kJ (Probabilistic) to 38.43 kJ (Chain), with growth evident but varying by structure. Between 10 and 20 services, energy growth plateaus, with medians clustering narrowly around 38-39 kJ (e.g., 38.68 kJ Probabilistic, 39.13 kJ Mesh), indicating steady-state energy usage under fixed workloads.
Median CPU utilization follows patterns consistent with energy trends. At 5 services, Parallel Fan-Out shows higher utilization at 1.87 cores, while Mesh remains lower at 0.88 cores. As system size grows to 10 services, values range from 1.03 cores (Probabilistic) to 1.72 cores (Chain); by 20 services, they converge around 0.9-1.0 cores across topologies (e.g., 0.97 cores for Seq FO, Chain, Hierarchical, Mesh), showing resource saturation reduces structural differences.
At 5 services, the Parallel Fan-Out topology achieves the highest median throughput (24.67 rps), followed closely by Sequential Fan-Out (22.75 rps) and Chain structures (24.55 rps). In contrast, the Mesh topology exhibits substantially lower throughput, indicating reduced request-handling capacity even at limited scale.
As system size increases to 10 services, throughput decreases across all topologies. Centralized structures maintain comparatively higher throughput, while the Mesh topology continues to show the lowest median value. At 20 services, values converge around 8-9 rps, such as 8.82 rps for Hierarchical, 8.79 rps for Chain, and 8.82 rps for Mesh, suggesting that resource saturation increasingly dominates structural effects at larger scale.
Response time trends mirror throughput inversely. At 5 services, Probabilistic topology shows the lowest median response time of 2.24 s, while Parallel Fan-Out achieves 2.05 s and Mesh exhibits higher response time at 13.43 s. As system size increases, median response time grows substantially for all topologies, particularly for structures with higher fan-out or dense connectivity.  In fact, Probabilistic increases at 2.24 s, Sequential Fan-Out at 3.81 s, and Mesh at 9.51 s. At 20 services, response time differences narrow across architectures converging around 9-10 s (e.g., 9.24 s for Mesh, 9.26 s for Chain and Hierarchical, 10.29 s for Parallel Fan-Out), indicating that resource contention (e.g., queuing of requests) can overshadow purely structural effects.
The fraction of requests that failed, in \%, varies by topology and system size. At 5 services, Probabilistic shows the lowest median failure rate (0.22\%), whereas Mesh shows the highest (2.61\%). Sequential and Parallel Fan-Out remain low (0.37--0.44\% at 5 services) and stay at 0\% median at 20 services. At 10 services, Mesh and Chain exhibit higher medians (2.48\% and 0.69\% respectively). At 20 services, Chain shows the highest median failure rate (2.13\%), while Seq FO and Par FO remain at 0\%. The dense Mesh topology consistently shows elevated failure rates at small and medium sizes, aligning with its lower throughput and higher response time under load.

\begin{figure}
  \centering
  \includegraphics[width=\linewidth]{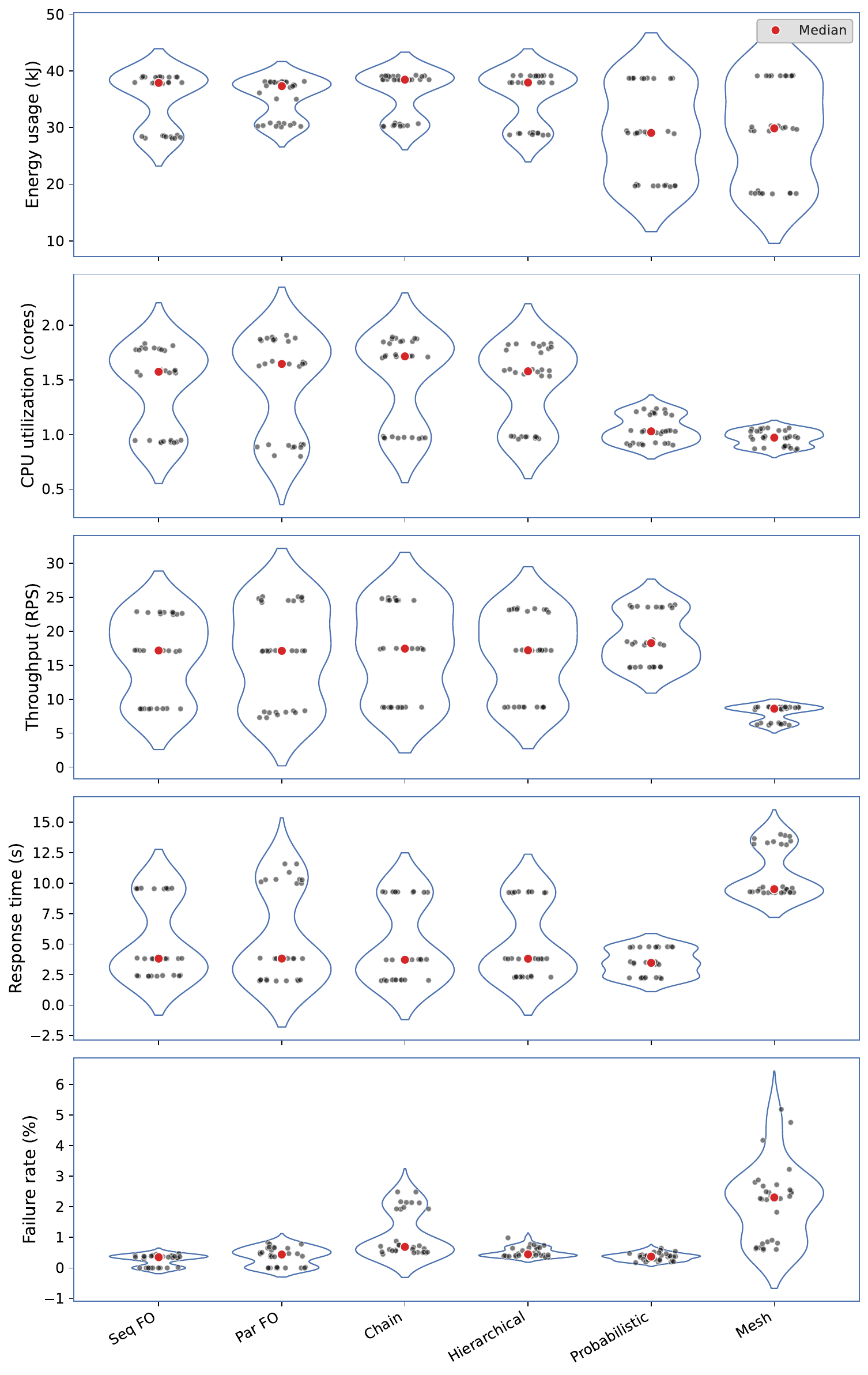}
  \caption{Metrics across topologies with sizes aggregated}
  \label{fig:boxplot_topologies}
  \vspace{-0.8cm}
\end{figure}

\textit{Comparison of Metrics by Topology.} Figure~\ref{fig:boxplot_topologies} illustrates the distribution of throughput, response time, energy usage, and CPU utilization across topologies, aggregated over system sizes. The violin plots visualize distribution density, while boxplots indicate interquartile range and median values.
For throughput, the Mesh topology exhibits a clearly lower distribution compared to all other structures, whereas Sequential and Parallel Fan-Out, Chain, and Hierarchical topologies achieve consistently higher values. Response time distributions show the inverse pattern, with the Mesh topology presenting substantially higher response times. These visual trends indicate a pronounced structural effect of dense connectivity on performance behavior.
Energy consumption shows comparatively smaller separation across topologies. While centralized and chain-like structures tend to exhibit slightly higher median energy values, the distributions overlap substantially. This suggests that architectural structure influences performance more strongly than energy demand under fixed workload conditions.
CPU utilization follows a pattern similar to energy consumption. Fan-out and chain structures exhibit higher utilization levels, whereas Mesh and Probabilistic topologies remain lower. Notably, the Mesh topology combines low throughput and high response time with comparatively lower CPU utilization, indicating that performance degradation in densely connected structures may stem from coordination overhead rather than sustained computational intensity.

\textit{Hypothesis Testing and Effect Size.}
Normality is assessed using Shapiro-Wilk tests per metric and topology. The majority of groups yielded $p < 0.05$, indicating violation of normality assumptions. Accordingly, non-parametric tests are used throughout: Kruskal-Wallis for global effects, Holm-corrected Mann--Whitney $U$ tests for structural contrasts, and Spearman rank correlation for performance--energy relationships. The significance level is $\alpha = 0.05$.
%
Kruskal-Wallis tests indicate statistically significant effects of topology on energy ($p = 0.001$), throughput, and response time ($p < 0.001$). The same applies to the system size factor and the topology size interaction ($p < 0.001$).

\begin{table}
\centering
\scriptsize
\caption{Contrasts between topology families. Bold indicates $p < 0.05$.}
\label{tab:structural_effects}
\setlength\tabcolsep{1pt}
\renewcommand\arraystretch{1.03}
\begin{tabular}{lcccc}
\toprule
Metric 
& Dense vs Others 
& Centralized vs Structured 
& Seq vs Par 
& Prob. vs Det. \\
\midrule
Energy Usage (kJ) & 0.236 & \textbf{0.024} & 0.569 & \textbf{0.003} \\\hdashline
CPU Usage & \textbf{1.10e-05} & 0.360 & 0.464 & \textbf{5.90e-05} \\\hdashline
Throughput (rps) & \textbf{8.35e-11} & 0.092 & 0.673 & 0.121 \\\hdashline
Response Time (s) & \textbf{9.03e-11} & 0.126 & 0.853 & 0.121 \\\hdashline
Failure rate & \textbf{1.49e-13} & \textbf{1.34e-09} & \textbf{0.019} & \textbf{0.014} \\
\bottomrule
\end{tabular}
\vspace{-0.8cm}
\end{table}

\textit{Structural Group Contrasts.} To complement per-topology descriptive analysis, statistical contrasts are also performed by aggregating topologies into broader structural families. We aggregate the six topologies into four structural families: \textbf{Centralized} (Sequential Fan-Out and Parallel Fan-Out), \textbf{Structured} (Chain with Branching and Hierarchical Tree), \textbf{Probabilistic} (Probabilistic Tree only), and \textbf{Dense} (Complex Mesh only). Centralized topologies share a single entry point that invokes downstream services (sequentially or in parallel). Structured topologies share a layered or chain-like dependency graph with deterministic invocation. Probabilistic is kept separate because invocation is stochastic; Dense is kept separate because of high connectivity and shared backends. This grouping allows contrasts such as Dense vs.\ others (coordination density), Centralized vs.\ Structured (orchestration vs.\ depth), Sequential vs.\ Parallel within Centralized (execution order), and Probabilistic vs.\ Deterministic (conditional execution). This aggregation enables evaluation of whether shared architectural characteristics systematically influence performance and energy behavior beyond individual topology instances. Because several topologies share common coordination patterns and dependency structures, family-level comparison provides a more generalizable view of structural effects. Holm-corrected contrasts 
(Table~\ref{tab:structural_effects}) indicate that densely connected structures differ significantly from more hierarchical and fan-out designs in throughput, response time, CPU utilization, and failure rate. For total energy (kJ), significant differences arise between some structural families (e.g., Centralized vs.\ Structured, Probabilistic vs.\ Deterministic), whereas Dense vs.\ others does not differ significantly at the group level. Probabilistic routing differs from deterministic structures primarily in energy and variability characteristics, while maintaining comparable throughput and response time distributions. Normalized efficiency metrics such as energy per request or per unit throughput reveal clearer structural differences than raw kJ totals, particularly for densely connected designs. Sequential and parallel centralized variants exhibit limited differences, suggesting that parallel fan-out alone does not fundamentally alter energy behavior under the evaluated workload.

\begin{table}
\centering
\scriptsize
\caption{Spearman correlation (performance vs.\ energy) per topology. Values shown as $\rho$ ($p$-value). Bold indicates $p < 0.05$.}
\label{tab:performance_energy}
\setlength\tabcolsep{4pt}
\renewcommand\arraystretch{1.03}
\begin{tabular}{lccc}
\toprule
Topology & Throughput--Energy & Resp.Time--Energy & Failure--Energy \\
\midrule
Probabilistic Tree 
& \textbf{-0.38 (0.038)} 
& \textbf{0.39 (0.035)} 
& 0.34 (0.066)  \\\hdashline

Parallel Fanout 
& \textbf{-0.38 (0.037)} 
& \textbf{0.38 (0.039)} 
& \textbf{0.43 (0.019)}  \\\hdashline

Chain With Branching 
& \textbf{-0.84 (6.78e-09)} 
& \textbf{0.84 (7.55e-09)} 
& \textbf{0.88 (9.64e-11)}  \\\hdashline

Complex Mesh 
& \textbf{0.96 (1.46e-16)} 
& \textbf{-0.96 (7.07e-17)} 
& \textbf{-0.72 (6.57e-06)}  \\\hdashline

Hierarchical Tree 
& \textbf{-0.84 (4.37e-09)} 
& \textbf{0.83 (1.27e-08)} 
& \textbf{0.80 (1.28e-07)}  \\\hdashline

Sequential Fanout 
& \textbf{-0.86 (1.36e-09)} 
& \textbf{0.85 (2.11e-09)} 
& \textbf{-0.64 (1.43e-04)}  \\
\bottomrule
\end{tabular}%
\vspace{-0.5cm}
\end{table}

\textit{Performance–Energy Relationship (RQ2).} Spearman correlation analysis (Table~\ref{tab:performance_energy}) examines whether performance and energy metrics co-vary across structural patterns. Results reveal topology-dependent coupling between performance and energy consumption. Hierarchical and fan-out structures show consistent monotonic relationships, where higher throughput corresponds to lower energy consumption and higher response time corresponds to increased energy demand. 
In contrast, highly interconnected structures exhibit stronger or amplified correlations, reflecting less efficient scaling behavior driven by longer execution paths and increased coordination overhead. Probabilistic routing shows weaker correlations, indicating more stable performance--energy trade-offs across repetitions.
Taken together, these results confirm that communication topology significantly influences performance and reliability, and shows statistically significant differences in energy consumption for several structural families (RQ1), and that the relationship between performance and energy varies across structural families (RQ2).

%% file: 7.discussion.tex
\section{Discussion}
\label{sec:discussion}
Communication topology significantly affects energy efficiency and performance in microservices. Since workload, service logic, and deployment remain constant, the differences stem from the structure of the service interaction graph.

\textit{\textbf{RQ1 - What is the impact of the architectural topology on the energy efficiency of microservices?}}
The statistical analysis confirms that communication topology has a measurable effect on the resulting energy usage.
Contrasts reveal that densely connected architectures (Mesh) differ significantly from other groups.
While average energy usage does not always isolate mesh as significantly higher, performance metrics consistently expose its inefficiencies, namely lower throughput and response time.
This indicates that total energy demand alone may obscure inefficiencies that become visible when accounting for performance metrics.
Probabilistic topology stands out as the most efficient solution for most of the combination between metrics and size.
Centralized versus structured contrasts also show significant differences in total energy consumption, suggesting that coordination depth and execution organization influence cumulative CPU activity even when computational workload per service is identical.
Because the workload is CPU-bound, energy consumption primarily reflects cumulative execution time and coordination cost. The results therefore indicate that topology-induced differences in call-graph structure directly affect energy efficiency.

\textit{\textbf{RQ2 - How does performance correlate to energy usage across topologies?}}
Performance metrics exhibit strong effects.
Dense connectivity significantly reduces throughput and increases both average and response time compared to centralized and hierarchical structures. These differences are statistically significant and consistent across repetitions.
Correlation analysis further reveals a systematic relationship between performance and energy consumption. For fan-out and tree-based topologies, higher throughput correlates with lower energy consumption, and higher latency correlates with higher energy demand. This indicates that faster execution reduces cumulative CPU activity.
In contrast, the Complex Mesh topology exhibits reversed or amplified correlations, reflecting inefficient scaling behavior. Longer execution paths, higher coordination overhead, and increased synchronization likely contribute to this divergence.
These findings suggest that, under CPU-bound workloads, performance optimization and energy efficiency are aligned objectives. Architectural structures that reduce sequential dependencies and coordination depth achieve both improved throughput and improved energy efficiency.
%
Failure rate analysis shows that topology influences system robustness in addition to performance and energy efficiency. Dense and chain-like structures exhibit significantly higher failure rates compared to centralized and hierarchical designs. Longer execution paths and higher coordination complexity increase the probability of request-level instability.
These findings reinforce prior observations that call-graph structure and service dependency patterns strongly affect system bottlenecks and instability~\cite{vanderlaan2023,luo2021,bakhtin2025network}.


\textit{\textbf{Interaction effect between topology and system size}} Significant topology–size interaction effects are observed for both throughput and energy metrics. While structural differences are present at small system sizes, divergence increases as the number of services grows.
At larger system sizes, dense connectivity exhibits pronounced throughput degradation and response time inflation, whereas hierarchical and fan-out structures maintain comparatively stable performance.
Architectural complexity increases with scale, so microservice systems efficient when small can degrade as they grow. 

\textit{\textbf{Implications for software architects}} The empirical findings have direct implications for the design of microservices.
First, topology selection should be treated as an important architectural decision. Structural properties such as connectivity density, coordination depth, and branching behavior measurably influence performance, energy efficiency, and reliability.
Second, normalized energy metrics provide more actionable insights than aggregate energy alone. Evaluating energy per request or per throughput unit reveals structural inefficiencies that total energy consumption may conceal.
Third, topology effects intensify with scale. Systems designed without considering interaction complexity may encounter disproportionate efficiency degradation as service count increases.
Fourth, in CPU-bound microservice deployments, improving performance frequently yields energy benefits. Reducing coordination overhead and limiting sequential service dependencies improves both throughput and energy efficiency simultaneously.
These insights are particularly relevant for architects of cloud-native systems, API gateway patterns, and large-scale service ecosystems operating in data-center environments. Finally, topology selection should also consider system scale. For smaller systems, probabilistic branching structures can provide favorable energy behavior, while for larger systems, Parallel Fan-Out offers a more consistent performance–energy trade-off. In contrast, densely connected Mesh topologies should be avoided in CPU-bound synchronous settings due to their high coordination overhead and reduced efficiency.

%% file: 8.threats.tex
\section{Threats to Validity}
\label{sec:threats}

\textit{Internal Validity.}
All experiments were executed on a dedicated single-node Kubernetes cluster under identical conditions with randomized order, namespace isolation, and cooldowns. Docker/Minikube were not restarted between runs, and CPU scaling or throttling was uncontrolled, potentially introducing noise. Larger topologies occasionally showed pod readiness delays but completed successfully. These factors can add variability but are unlikely to explain topology effects.

\textit{External Validity.}
We study six synthetic topology families at three system sizes (5, 10, 20 services) under a CPU-bound workload on a single-node cluster. Real microservices combine I/O-bound, database, and network interactions. Thus, our results may not generalize to distributed systems. The single-node setup omits cross-node communication costs, and larger industrial deployments may show different quantitative behaviors and topology–energy trade-offs.

\textit{Construct Validity.}
Topologies are modeled with validated $\mu$Bench workmodels, capturing structural variations but not full architectural diversity such as asynchronous, event-driven, or hybrid systems. Energy is measured via Intel RAPL counters (CPU and DRAM only); network and storage energy are excluded, which may understate costs for dense Mesh topologies. Metrics include total and normalized energy values, abstracting per-service accounting.

\textit{Conclusion Validity.}
Non-normal data justify non-parametric analysis (Kruskal–Wallis, Mann–Whitney $U$ with Holm correction). Ten repetitions limit power for small effects, so we interpret only consistent, large differences. Topology–size interactions are assessed at small scales (5–20 services) and may differ in production. The results are robust within controlled CPU-bound settings but should be generalized cautiously.

%% file: 9.conclusion.tex
\section{Conclusions}\label{sec:conclusions}

This study demonstrates that communication topology substantially affects microservice system performance and energy use. Even under identical workloads, architectural structure drives statistically significant differences: dense topologies show lower throughput, higher response time, and greater energy consumption, while hierarchical and fan-out designs perform more efficiently. Topology–size interactions suggest these effects intensify as systems scale, highlighting the scale-dependent impact of architectural design on efficiency and robustness.
Future work should extend this analysis by generating controlled topology families (e.g., star, chain, tree, cyclic, layered, database-centric) using $\mu\text{Bench}$ and scaling experiments beyond single-node Minikube to large Kubernetes clusters (200+ services). Such studies would better capture the behavior of real industrial systems with hundreds or thousands of services.